# Sensitivity Analysis of Resonant Circuits

Olivier Buu

*Abstract* — We use first-order perturbation theory to provide a local linear relation between the circuit parameters and the poles of an RLC network. The sensitivity matrix, which defines this relationship, is obtained from the systems eigenvectors and the derivative of its eigenvalues. In general, the sensitivity matrix is related to the equilibrium fluctuations of the system. In particular, it may be used as the basis for a statistical model to efficiently predict the sensitivity of the circuit response to small component variations. The method is illustrated with a calculation of conditional probabilities by Monte Carlo Simulation.

## I. Introduction

SENSITIVITY analysis is an integral part of computer-aided circuit design. Efficient statistical analysis algorithms are available to simulate the circuit response at fixed frequencies [1], from which the sensitivity to components variation may be obtained by regression. In the context of resonant circuits, however, the designer is primarily interested in the poles and the circuit response on-resonance. Tracking the resonances at each trial requires an extra computational step that undermines the efficiency of existing methods.

In this paper, we follow the reverse procedure: we first determine the local linear relationship between the circuit parameters and the poles and response of the system, then carry out "primitive" Monte Carlo simulations. For yield predictions, which require a large number of trials, the fixed cost associated with the determination of the sensitivity matrix is lower than the recurring costs in existing approaches.

We define the sensitivity matrix as the Jacobian of the transformation between the circuit parameters and the poles and system response [2]. We present a general method to calculate the sensitivity matrix based on the solution of the eigenvalue problem associated with the circuit. This method is illustrated with a simple example.

## II. Background

Consider the transfer function $H$ of an arbitrary network of resistances, inductors, and capacitors. The observable response of the circuit is entirely characterized by the poles $s_1, \ldots, s_N$ and the values of the transfer function on-resonance $H_1, \ldots, H_N$[1]. Formally, there exists a complicated relationship between the circuit parameters $\boldsymbol{h} = [h_1, \ldots, h_K]$ and the column vector $\boldsymbol{\Omega}$ formed by the real and imaginary parts of the poles and each independent component of the transfer function at each resonance frequency. To study the sensitivity to small component variations $\boldsymbol{\delta h}$, we follow [2] and linearize this relation:

$$\boldsymbol{\delta\Omega} = [\nabla_h \boldsymbol{\Omega}]^T . \boldsymbol{\delta h} \tag{1}$$

where $[\ ]^T$ denotes the matrix transpose. In practice, only a subset of the observable parameters may be under specification, and the size of the sensitivity matrix is reduced accordingly.

Although the calculation of the Jacobian $\nabla_h \boldsymbol{\Omega}$ may be computationally costly for large systems, the simple linear relation (1) allows for efficient statistical analysis of the circuit. In some cases, the Jacobian gives direct access to the multivariate distribution function. In particular, if the relation between $\boldsymbol{h}$ and $\boldsymbol{\Omega}$ is bijective, the probability density function $g$ associated with $\boldsymbol{\Omega}$ is known locally from the relation:

$$g(\boldsymbol{\Omega}) = f(\boldsymbol{h}) |\nabla_h \boldsymbol{\Omega}|^{-1} \tag{2}$$

where $f$ is the probability density function associated with the random vector $\boldsymbol{h}$. More generally, when the second moment of the distribution of the circuit parameters exists, the co-variance matrix for the random vector $\boldsymbol{\Omega}$ is given by:

$$\Sigma_\Omega = [\nabla_h \boldsymbol{\Omega}] \Sigma_h [\nabla_h \boldsymbol{\Omega}]^T \tag{3}$$

where $\Sigma_h$ is the co-variance matrix of the vector $\boldsymbol{h}$.

## III. Sensitivity Matrix Calculation

The sensitivity matrix is assembled from the derivatives of the poles and the transfer function with respect to the circuit parameters. In this section, we review the perturbation method used to calculate these derivative terms based on the solution of the circuit eigenvalue problem. The eigenvalue problem is formulated from the circuit state equation.

### A. State Equation

The circuit equation for a Linear Time-Invariant network is assumed to take the standard form:

$$M\dot{x} = -Nx + Bu \tag{4}$$

---

[1] This list may include a pole at infinity.

where the vector $x$ includes the node voltages at the capacitors terminals and the currents flowing through the inductors. $B$ is the input matrix and $u$ is the excitation vector. $M$ is a matrix composed of capacitance and inductance values. The matrix $N$ includes the resistance values and the node-branch incidence matrix describing the network under study. Both $M$ and $N$ may be written as sums of sparse matrices corresponding to the individual component contributions. These sparse matrices, sometimes called "component stamps" in the literature [3], are useful for the sensitivity analysis presented below. Provided $x$ does not contain any linearly-dependent variables, then (4) is a state equation and the matrix $M$ is full-rank. We will assume this condition to be fulfilled in the rest of this paper.

The transfer function is determined by the output equation:

$$v = Dx + Eu \qquad (5)$$

where $v$ is the output vector, $D$ is the output matrix, and $E$ is the transmission matrix. By taking the Laplace transform of (3) and (4) and applying the definition of the transfer function to the zero-state output vector, we obtain:

$$H(s) = D(N + sM)^{-1}B + E \qquad (6)$$

where $s$ is the Laplace variable.

### B. Generalized eigenvalue problem

The circuit is entirely characterized by the eigenvalues and eigenvectors of the state equation. The square matrices of right eigenvectors $X = [x_1, ..., x_N]$ and left eigenvectors $Y = [y_1, ..., y_N]$ are solutions of the generalized eigenvalue problem:

$$-NX = MX\Lambda \quad \text{and} \quad -Y^T N = \Lambda Y^T M \qquad (7)$$

where $\Lambda = \text{diag}[s_1, ..., s_N]$ is a diagonal matrix of eigenvalues. Since $M$ and $N$ are real matrices, the eigenvalues are either real or complex conjugate pairs. For passive networks, all the eigenvalues are located in the left half of the complex plane.

From (7), it can be shown that the eigenvectors are bi-orthogonal. Since $M$ is assumed to be non-singular, we can always find a normalization such that:

$$-Y^T MX = I \quad \text{and} \quad -Y^T NX = \Lambda \qquad (8)$$

where $I$ is the identity matrix. Using (5) and (7) we obtain the following expression for the transfer function:

$$H(s) = DX(\Lambda - sI)^{-1}Y^T B + E \qquad (9)$$

Together with the eigenvalues, this last expression forms the basis of the sensitivity analysis described in the next sections.

### C. Derivative of Eigenvalues

For simple eigenvalues, the derivatives with respect to a circuit parameter $h_k$ is [4]:

$$\frac{\partial \Lambda}{\partial h_k} = \text{diag}\left[Y^T \frac{\partial N}{\partial h_k} X\right] + \Lambda \, \text{diag}\left[Y^T \frac{\partial M}{\partial h_k} X\right] \qquad (10)$$

As noted above, the matrix derivatives $\frac{\partial M}{\partial h_k}$ and $\frac{\partial N}{\partial h_k}$ are sparse and closely related to the "stamp" for the circuit component parameterized by $h_k$.

The case of multiple eigenvalues is addressed in [5]: for an eigenvalue $s_p$ of multiplicity $P > 1$ with associated eigenvectors $X_p = [x_{p1}, ..., x_{pP}]$ and $Y_p = [y_{p1}, ..., y_{pP}]$, there are $P$ derivatives which are the eigenvalues of the matrix $Y_p^T \left[\frac{\partial N}{\partial h_k} + s_p \frac{\partial M}{\partial h_k}\right] X_p$.

### D. Derivative of the Transfer Function

The derivative of the on-resonance transfer function includes two terms:

$$\frac{\partial H_n}{\partial h_k} = \left[\frac{\partial H}{\partial h_k}\right] + j\left[\frac{\partial \omega_n}{\partial h_k}\right]\left[\frac{\partial H}{\partial s}\right]_{s=j\omega_n} \qquad (11)$$

where $\omega_n$ is the imaginary part of the n[th] pole. To obtain these two terms we introduce some intermediate calculation steps:

$$Z_n = (N + j\omega_n M)^{-1} = X(\Lambda - j\omega_n I)^{-1} Y^T \qquad (12)$$
$$B_n = Z_n B \qquad (13)$$
$$D_n = DZ_n \qquad (14)$$

In the previous expressions, we have assumed that the resonance of interest is damped, so the matrix $(N + j\omega_n M)$ is non-singular. Note that (12) does not require a full matrix inversion. The right-hand-side terms of (11) follows from derivatives of (6):

$$\left[\frac{\partial H}{\partial s}\right]_{s=j\omega_n} = -D_n M B_n \qquad (15)$$

$$\left[\frac{\partial H}{\partial h_k}\right] = \frac{\partial D}{\partial h_k} B_n + D_n \frac{\partial B}{\partial h_k} + \frac{\partial E}{\partial h_k}$$
$$- D_n \left[\frac{\partial N}{\partial h_k} + s_n \frac{\partial M}{\partial h_k}\right] B_n \qquad (16)$$

## IV. COMPUTATIONAL COST

The most costly step of the sensitivity matrix calculation is the solution of the eigenvalue problem, which scales as $\mathcal{O}(N^3)$ operations. If the probability density function can be obtained, from equ. (2) or otherwise, the cost of a Monte Carlo trial is the cost of sampling the distribution. In the worst case, an additional matrix multiplication (equ. (1)) is required, $\mathcal{O}(N^2)$.

By contrast, the cost of setting up a quadratic approximation of the response function at a given frequency is $\mathcal{O}(K)$. Since the number of degrees of freedom is equal the number of independent active elements, this cost is equivalent to $\mathcal{O}(N)$. Each simulation involves a matrix multiplication, with a $\mathcal{O}(N^2)$ cost. However, the response function has to be evaluated at $N$ different frequencies to characterize the circuit, so the fixed and recurring costs are, respectively $\mathcal{O}(N^2)$ and $\mathcal{O}(N^3)$ in this method. Moreover, there is a cost associated with tracking the resonance frequencies at each trial.

The standard deviation of yield predictions scales as the inverse of the square root of the number of trials. For a ~$10^{-2}$ accuracy on simulation results, we assume a $10^4$ trials run. Considering a circuit with $N = 50$ eigenmodes, the sensitivity matrix method would require ~$10^7$ operations. The same simulation would cost ~$10^9$ operations to achieve the same accuracy with the quadratic approximation.

## V. EXAMPLE

Fig. 1 shows a circuit example used in Magnetic Resonance instruments [6]. The inductive transducer $L_{coil}$ is embedded in a two-port matching network, where port 1 is tuned to 200 MHz, and port 2 to 50 MHz.

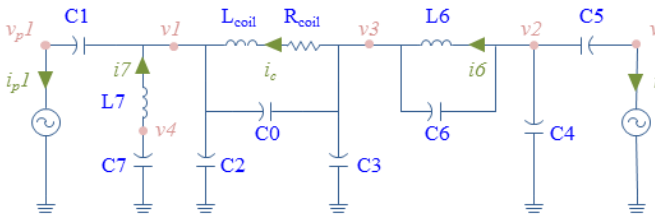

Fig. 1. Two-port matching network. Component values are $C0 = 3.47$ pF, $C1 = 0.68$ pF, $C2 = 2.28$ pF, $C3 = 1.84$ pF, $C4 = 42.08$ pF, $C5 = 4.825$ pF, $C6 = 12.7$ pF, $C7 = 25.33$ pF, $L6 = 50$ nH, $L7 = 400$ nH, $L_{coil} = 150$ nH, and $R_{coil} = 0.47$ Ohm.

In this RF application, the S-parameters are the preferred signal representation. However, it is convenient to first obtain the 2x2 port-admittance matrix and component stamps by Modified Nodal Analysis. In this case, the matrices $\boldsymbol{M}$ and $\boldsymbol{N}$ are:

$$\boldsymbol{M} = \begin{bmatrix} C & 0 \\ 0 & L \end{bmatrix} \text{ and } \boldsymbol{N} = \begin{bmatrix} 0 & A \\ -A^T & R \end{bmatrix} \quad (17)$$

The sub-matrices are obtained by inspection:

$$\boldsymbol{C} = \begin{bmatrix} C_0 + C_1 + C_2 & 0 & -C_0 & 0 \\ 0 & C_4 + C_5 + C_6 & -C_6 & 0 \\ -C_0 & -C_6 & C_0 + C_3 + C_6 & 0 \\ 0 & 0 & 0 & C_7 \end{bmatrix} \quad (18)$$

$$\boldsymbol{L} = \text{diag}[L_6, L_7, L_{coil}] \quad (19)$$

$$\boldsymbol{R} = \text{diag}[0, 0, R_{coil}] \quad (20)$$

$$\boldsymbol{A} = \begin{bmatrix} 0 & 1 & 1 \\ -1 & 0 & 0 \\ 1 & 0 & -1 \\ 0 & -1 & 0 \end{bmatrix} \quad (21)$$

We used Matlab™ to generate the eigenvalues and eigenvectors, which are reported in table 1. In addition to the nominal resonances at 200 MHz and 50 MHz, there is a spurious resonance at 179 MHz and a pole at DC.

TABLE I
EIGENVALUES AND EIGENVECTORS

| | Channel 1[a] | Channel 2[a] | Spurious[a] | DC |
|---|---|---|---|---|
| Frequency[b] | 199.9 MHz | 50.0 MHz | 179.0 MHz | 0 |
| $v1$[c] | -1.0 | 0.0 | -1.0 | 1 |
| $v2$[c] | 0.0 | -0.5 | 0.0 | 1 |
| $v3$[c] | 0.7 | -0.4 | -0.5 | 1 |
| $v4$[c] | 0.1 | 1.0 | 0.1 | 1 |
| $i6$[d] | -0.1-j10.8 | -j8.2 | j10.4 | 0 |
| $i7$[d] | j2.1 | j7.9 | j2.4 | 0 |
| $i_{coil}$[d] | j8.9 | -j8.4 | j2.7 | 0 |

[a]Right eigenvector from the eigenvalue of positive natural frequency.
[b]Each frequency corresponds to a complex-conjugate pair of poles.
[c]Voltage nodes in V.
[d]Inductor currents in mA.

Because we eliminated the redundant variables $i_p1, v_p1, i_p2, v_p2$ from the state vector, the driving term depends on the time derivative on the input vector [7] and takes the form $\boldsymbol{B\dot{u}}$ with:

$$\boldsymbol{B} = -\begin{bmatrix} C_1 & 0 \\ 0 & C_5 \\ 0 & 0 \\ 0 & 0 \end{bmatrix} \text{ and } \boldsymbol{u} = \begin{bmatrix} v_p 1 \\ v_p 2 \end{bmatrix} \quad (22)$$

Similarly, the transmission term in the output equation takes the form $\boldsymbol{E\dot{u}}$ because the DC mode is not observable [7]. The coefficients of the output equation are:

$$\boldsymbol{D} = -\boldsymbol{B}^T \text{ and } \boldsymbol{v} = \begin{bmatrix} i_p 1 \\ i_p 2 \end{bmatrix} \quad (23)$$

$$\boldsymbol{E} = \begin{bmatrix} C_1 & 0 \\ 0 & C_5 \end{bmatrix} \quad (24)$$

These modifications to the standard circuit equation add terms to (15) but do not drastically alter the sensitivity analysis. The scattering matrix $\boldsymbol{S}(s)$ follows from the well-known relation:

$$\boldsymbol{S} = (\boldsymbol{I} - Z_0\boldsymbol{H})(\boldsymbol{I} + Z_0\boldsymbol{H})^{-1} \quad (25)$$

where $Z_0 = 50$ Ohm is the characteristic port impedance. Differentiating (25) yields:

$$\partial \boldsymbol{S} = \tfrac{1}{2}(\boldsymbol{I} + \boldsymbol{S})Z_0 \partial \boldsymbol{H}(\boldsymbol{I} + \boldsymbol{S}) \quad (26)$$

In this example the specifications are:

$$|\omega_1 - 200 \text{ MHz}| < 5 \text{ MHz} \quad (27)$$
$$|\omega_2 - 50 \text{ MHz}| < 5 \text{ MHz} \quad (28)$$
$$|S_{11}(j\omega_1)| < -10 \text{ dB} \quad (29)$$
$$|S_{22}(j\omega_2)| < -10 \text{ dB} \quad (30)$$

Correspondingly, we characterize the circuit response with the vector:

$$\mathbf{\Omega} = [\omega_1, \omega_2, \Re(S_{11}), \Im(S_{11}), \Re(S_{22}), \Im(S_{22})]^T \quad (31)$$

where $\Re$ denotes the real part and $\Im$ the imaginary part. The corresponding sensitivity matrix is compiled in table 2.

TABLE II
SENSITIVITY MATRIX

| Component | $\frac{\partial \omega_1}{\partial h}$ | $\left|\frac{\partial S_{11}}{\partial h}\right|$ | $\frac{\partial \omega_2}{\partial h}$ | $\left|\frac{\partial S_{22}}{\partial h}\right|$ |
|---|---|---|---|---|
| $C_0$ | -14.4[a] | 1.78[d] | -0.1[a] | 0.06[d] |
| $C_1$ | -5.1[a] | 3.85[d] | 0.0[a] | 0.00[d] |
| $C_2$ | -5.1[a] | 1.87[d] | 0.0[a] | 0.00[d] |
| $C_3$ | -2.4[a] | 0.55[d] | -0.1[a] | 0.06[d] |
| $C_4$ | 0.0[a] | 0.00[d] | -0.2[a] | 0.08[d] |
| $C_5$ | 0.0[a] | 0.00[d] | -0.2[a] | 0.71[d] |
| $C_6$ | -2.4[a] | 0.55[d] | 0.0[a] | 0.00[d] |
| $C_7$ | 0.0[a] | 0.01[d] | -0.6[a] | 0.00[d] |
| $L_6$ | -0.6[b] | 0.14[e] | 0.0[b] | 0.00[e] |
| $L_7$ | 0.0[b] | 0.01[e] | 0.0[b] | 0.00[e] |
| $L_{coil}$ | -0.4[b] | 0.02[e] | 0.0[b] | 0.00[e] |
| $R_{coil}$ | 0.0[c] | 4.14[f] | 0.0[c] | 4.04[f] |

[a] In units of MHz/pF.
[b] In units of MHz/nH.
[c] In units of MHz/Ohm.
[d] In units of 1/pF.
[e] In units of 1/nH.
[f] In units of 1/Ohm.

Assuming a 5% variance and no correlation between the component values, we can calculate the second moment of the chosen engineering parameters from the diagonal elements of the co-variance matrix $\mathbf{\Sigma_\Omega}$ obtained from (3). The values are listed in table 3.

TABLE III
PREDICTED YIELD

| | $\omega_1$[a] | $\|S_{11}\|$[b] | $\omega_2$ | $\|S_{22}\|$ |
|---|---|---|---|---|
| Variance | 4.54 MHz | 0.69 | 1.25 MHz | 0.25 |
| Partial Yields | 73% | 35% | 79% | 79% |
| Combined Yields | 26% | | 62% | |
| Total Yield | 19% | | | |

[a] $\sqrt{\mathbb{E}(\omega_1 - 200 \text{ MHz})^2} = \sqrt{\mathbf{\Sigma_\Omega}_{11}}$

[b] $\sqrt{\mathbb{E}|S_{11}|^2} = \sqrt{\mathbf{\Sigma_\Omega}_{33} + \mathbf{\Sigma_\Omega}_{44}}$

To calculate the yield, we further assume the circuit parameters to be normally distributed. The distribution of the random vector $\mathbf{\Omega}$ is then multivariate normal. Instead of integrating the 6-dimensional probability density function over the specification domain, we found it more accurate and faster to use a Matlab routine to sample the distribution. The various conditional probabilities, estimated by averaging $10^9$ trials, are reported in table 3. The coefficient of variation on these figures is ~$10^{-3}$. A convenient representation of the results is the area-proportional Venn diagram, which was created with the `VennEuler` algorithm [10] and is shown on Fig. 2.

In this example, the detailed yield analysis points to the return loss on channel 1 as the largest risk of failure. The sensitivity matrix suggests that reducing the variance of $C_1$, $C_0$, $C_2$ and $R_{coil}$ would improve the yield. Replacing $C_1$ by a trimmer capacitor and adding a tuning process is another solution to the yield issue. These different assumptions and their economic implications may be tested by re-calculating the yield with the Monte Carlo method. The experimental validation of the model may be done by Design of Experiment (DOE) based on the sensitivity matrix calculated above.

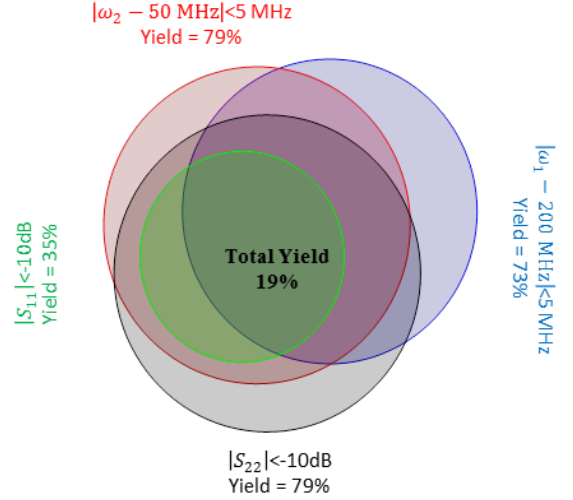

Fig. 2: Area-proportional Venn diagram of the Monte-Carlo Simulation. The surface area of each disk is proportional to the corresponding population. The overlaps between different disks show the conditional probabilities. The region overlapped by all four disks represents the total yield.

## VI. CONCLUSION

In this paper we have shown that the sensitivity matrix method is an efficient way to carry out the statistical analysis of resonant circuits. Despite the computational cost of determining the sensitivity matrix, its value may be realized in the direct calculation of the co-variance matrix, DOE studies, or substantiating causal analyses used in "six-sigma" quality control frameworks.

Our approach may be improved with more efficient eigenvectors calculation algorithms [8] or faster Monte Carlo methods [9]. A low-cost approximation of the sensitivity matrix may also be obtained by the quadratic approximation, if the eigen-frequencies are known in advance. Finally, this approach is applicable to any linear time-invariant system, and may be expanded to other network characterizations, as described in the appendix.

## VII. ACKNOWLEDGMENTS

This work is an offshoot of Dr. M. A. Smith's Six-Sigma green belt project. We also thank Bob Taber for drawing our attention to the importance of eigenmodes in resonant circuits.

APPENDIX

ALTERNATE CIRCUIT CHARACTERIZATIONS

Engineering specifications may include circuit characterizations other than the quantities we considered in the analysis presented above. In this appendix, we provide the elements of perturbation theory that may be used with other well-known characterizations. For simplicity, we restrict this section to the case of circuits with single eigenvalues. The case of multiple eigenvalues has been worked out [10] but is outside the scope of this paper.

*A. Time Domain Analysis*

The circuit natural response may be calculated in terms of eigenvectors from (4) and (7):

$$\boldsymbol{x}(t) = \boldsymbol{X}e^{\Lambda t}\boldsymbol{Y}^T\boldsymbol{M}\boldsymbol{x}(0) \tag{32}$$

where $\boldsymbol{x}(0)$ is the vector of initial conditions. Since $\boldsymbol{\Lambda}$ is diagonal, the calculation of the exponential term presents no numerical difficulty. The differentiation of (30) with respect to a circuit parameter $h_k$ involves the derivative of eigenvectors. Ref. [11] gives their expression as linear combinations of the un-perturbed eigenvectors in the case of distinct eigenvalues:

$$\frac{\partial \boldsymbol{x}_n}{\partial h_k} = \frac{1}{2}\boldsymbol{y}_n^T \frac{\partial \boldsymbol{M}}{\partial h_k}\boldsymbol{x}_n - \sum_{m \neq n} \frac{\boldsymbol{y}_m^T\left[\frac{\partial \boldsymbol{N}}{\partial h_k} + s_n\frac{\partial \boldsymbol{M}}{\partial h_k}\right]\boldsymbol{x}_n}{s_n - s_m}\boldsymbol{x}_m \tag{33}$$

An alternate expression involving only one un-perturbed eigenvector is given by [12].

*B. Residues*

The matrix $\boldsymbol{H}(s)$ may be expanded as a sum of rational functions:

$$\boldsymbol{H}(s) = \sum_{n=1}^{N}\frac{\boldsymbol{K}_n}{s-s_n} + \boldsymbol{E} \tag{34}$$

The poles $s_n$ and residue matrices $\boldsymbol{K}_n$ provide a complete characterization of the observable response. By expanding (8) we can express the residue matrices in terms of eigenvectors:

$$\boldsymbol{K}_n = (\boldsymbol{D}\boldsymbol{x}_n)(\boldsymbol{y}_n^T\boldsymbol{B}) \tag{35}$$

Similarly to the time-domain analysis, the perturbation of the residue matrices involves the derivative of eigenvectors.

*C. Zeros*

Single Input Single Output systems are often analyzed in terms of pole-zero loci. The first-order derivative of a zero $z_p$ may be obtained by differentiating the implicit relation $H(z_p)=0$ with respect to the circuit parameter $h_k$:

$$\frac{\partial z_p}{\partial h_k} = -\frac{\left[\frac{\partial H}{\partial h_k}\right]_{s=z_p}}{\left[\frac{\partial H}{\partial s}\right]_{s=z_p}} \tag{36}$$

The numerator and denominator are obtained similarly to (15) and (16). In this case, the vectors $\boldsymbol{B}_n$ and $\boldsymbol{D}_n$ may be interpreted as the respective solutions of the direct and adjoint systems at $s = z_p$.